\documentclass[a4paper]{jpconf}
\usepackage{graphicx}
\usepackage{cite}
\begin{document}
\title{Event-based simulation of neutron experiments: interference, entanglement and uncertainty relations }

\author{Kristel Michielsen$^{1,2}$ and Hans De Raedt$^3$}

\address{$^1$ Institute for Advanced Simulation, J\"ulich Supercomputing Centre, \\Forschungszentrum J\"ulich, D-52425 J\"ulich, Germany}
\address{$^2$ RWTH Aachen University, D-52056 Aachen, Germany}
\address{$^3$ Department of Applied Physics, Zernike Institute for Advanced Materials, \\University of Groningen, Nijenborgh 4, NL-9747 AG Groningen, The Netherlands}

\ead{k.michielsen@fz-juelich.de, h.a.de.raedt@rug.nl}

\begin{abstract}
We discuss a discrete-event simulation approach, which has been shown to give a unified cause-and-effect description of many quantum optics and single-neutron interferometry experiments.
The event-based simulation algorithm does not require the knowledge of the solution of a wave equation of the whole system, yet reproduces the corresponding statistical distributions
by generating detection events one-by-one. It is showm that single-particle interference and entanglement, two important quantum phenomena, emerge via information exchange between
individual particles and devices such as beam splitters, polarizers and detectors.
We demonstrate this by reproducing the results of several single-neutron interferometry experiments, including
one that demonstrates interference and one that demonstrates the violation of a Bell-type inequality.
We also present event-based simulation results of a single neutron experiment designed to test the validity
of Ozawa's universally valid error-disturbance relation, an uncertainty relation derived using the theory of general quantum measurements.
\end{abstract}

\section{Introduction}
The mathematical framework of quantum theory allows for the calculation of numbers which can be compared with experimental
data as long as these numbers refer to statistical
averages of measured quantities.
However, as soon as an experiment records individual detector
clicks which contribute to the statistical average of a quantity then a fundamental problem, related to the quantum measurement problem~\cite{HOME97,BALL03}, appears.
An example of such experiments are interference experiments in which the interference
pattern is built up by successive discrete detection events.
Another example are Bell-test experiments in which the correlations between two degrees of freedom
are computed as averages of pairs of detection events which are
seen to take values that are reminiscent of those of the singlet state in the quantum theoretical description of the thought experiment.

An intriguing question to be answered is why individual entities which do not interact with each other can exhibit the collective behavior
that gives rise to the observed interference pattern and why two degrees of freedom
can show correlations corresponding to those of the singlet state.
Since quantum theory postulates that it is fundamentally impossible to go beyond the description in terms of probability distributions,
an answer in terms of a cause-and-effect description of the observed phenomena cannot be given within the framework of quantum
theory.
We provide an answer by constructing an event-based simulation model that reproduces the statistical distributions of
quantum (and Maxwell's) theory without solving a wave equation but by modeling physical phenomena as a chronological sequence of events
whereby events can be actions of an experimenter, particle emissions by a source, signal generations by a detector,
interactions of a particle with a material and so on~\cite{MICH11a,RAED12a,RAED12b}.
The underlying assumption of this simulation approach is that current scientific knowledge derives from the discrete events which
are observed in laboratory experiments and from relations between those events.
Hence, the event-based simulation approach concerns what we can say about these experiments but not what ``really'' happens in Nature.

\begin{figure}[t]
\begin{center}
\includegraphics[width=15cm]{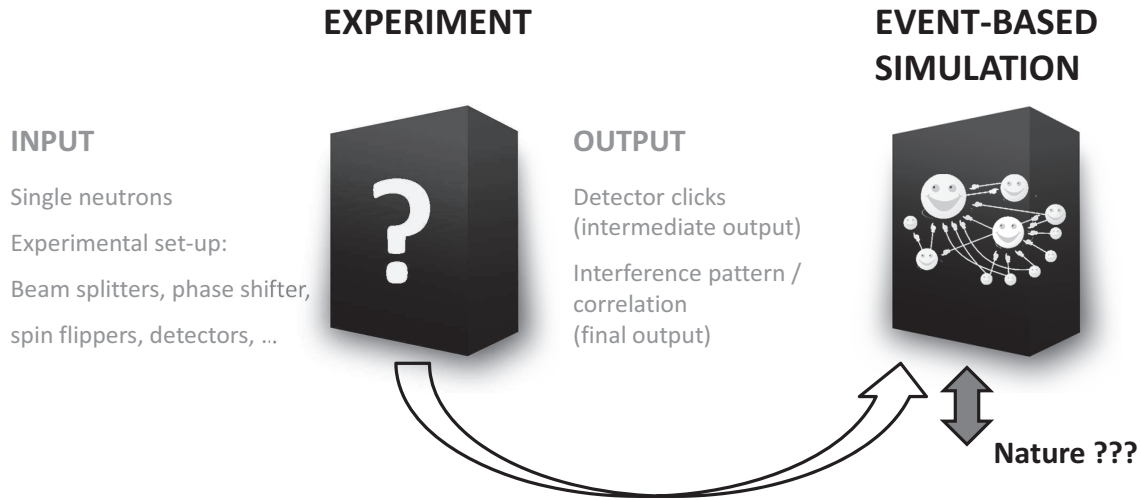}
\caption{%
Schematic of the working principle of the event-based simulation approach.
The first step consists of an analysis of the experiment, keeping in mind that
the devil is in the detail. Information about the input, such as characteristics of the particle source
and all other components in the experimental setup, and about the output  such as the detector clicks
(intermediate output) and the interference pattern or correlation (final output) including
the data analysis procedure, is collected.
It is assumed that no information is available about how the input is transformed into the output.
In a second step the ``black box'' that connects input and output in the experiment
is replaced by a set of simple rules that transform the same input into the same output.
The question whether the rules describe what is going in Nature cannot be answered since
information necessary to answer this question is not available.
}
\label{fig1}
\end{center}
\end{figure}

The general idea of the event-based simulation method (see Fig.~\ref{fig1}) is that simple rules define discrete-event processes which may lead to the
behavior that is observed in experiments. The basic strategy in designing these rules is to carefully examine
the experimental procedure and to devise rules such that they produce the same kind of data as those
recorded in experiment, while avoiding the trap of simulating thought experiments that are difficult to realize in
the laboratory. Evidently, mainly because of insufficient knowledge, the rules are not unique.
Hence, the simplest rules one could think of can be used until a new experiment indicates otherwise.
The method may be considered to be entirely classical since it only uses concepts of the macroscopic world,
but some of the rules are not those of classical Newtonian dynamics.
An overview of the method and its applications can be found in Refs.~\cite{MICH11a,RAED12a,RAED12b}.
Here we present an application of the method to single-neutron experiments showing
interference~\cite{RAUC74a,KROU00}, the violation of a Bell-type inequality~\cite{HASE03}, the validity
of Ozawa's universally valid error-disturbance relation~\cite{ERHA12,SULY13}.

\section{Neutron interference}
One of the most fundamental experiments in quantum physics is the single-particle double-slit experiment.
Feynman stated that the phenomenon of electron diffraction by a double-slit structure is ``impossible,
{\sl absolutely} impossible, to explain in any classical way, and has in it the heart of quantum mechanics.
In reality it contains the only mystery.''~\cite{FEYN65}
While Young's original double-slit experiment helped to
establish the wave theory of light~\cite{YOUN02}, variants of the experiment over the years with electrons~\cite{DONA73,MERL76,TONO89,FRAB08,HASS10,ROSA12,FRAB12,BACH13},
single photons~\cite{JACQ05,SAVE02,GARC02,KOLE13},
neutrons~\cite{RAUC74a,ZEIL88,RAUC00}, atoms~\cite{KEIT91,CARN91}, molecules~\cite{ARND99,BREZ02,JUFF12} and droplets~\cite{COUD06}
provided illustrations of the necessity of wave-particle duality in quantum theory~\cite{BALL03}.

In classical optics, both diffraction and interference play a role in the two-slit experiment
and are manifestations of the wave character of these phenomena.
However, an interference pattern identical in form to that of classical optics can be
observed by collecting many detector spots or clicks which are the result of electrons, photons, neutrons, atoms or molecules travelling one-by-one through a double-slit structure.
In these experiments the so-called interference pattern is the statistical distribution of the detection events (spots at or clicks of the detector).
Hence, in these particle-like experiments, only the correlations between detection events reveal interference.
In what follows we use the term interference pattern for the statistical distribution of detection events.

Various experiments have been designed to study interference phenomena.
In simple terms, an interferometer is a device containing a beam splitter that splits an incident beam into two or more mutually coherent beams.
These beams are then recombined, usuually by employing mirrors and/or additional beam splitters, on a detection screen showing interference fringes.
A well-known type of interferometer which is frequently used is the Mach-Zehnder interferometer (MZI).
In a MZI the incoming beam is split by a beam splitter. The two resulting beams are each reflected by a mirror.
One of the beams passes through a phase shifter, then both beams pass through a second beam splitter and  impinge on two detectors which are placed behind the two output beams of the second beam splitter.

\begin{figure}[pt]
\begin{center}
\includegraphics[width=8cm]{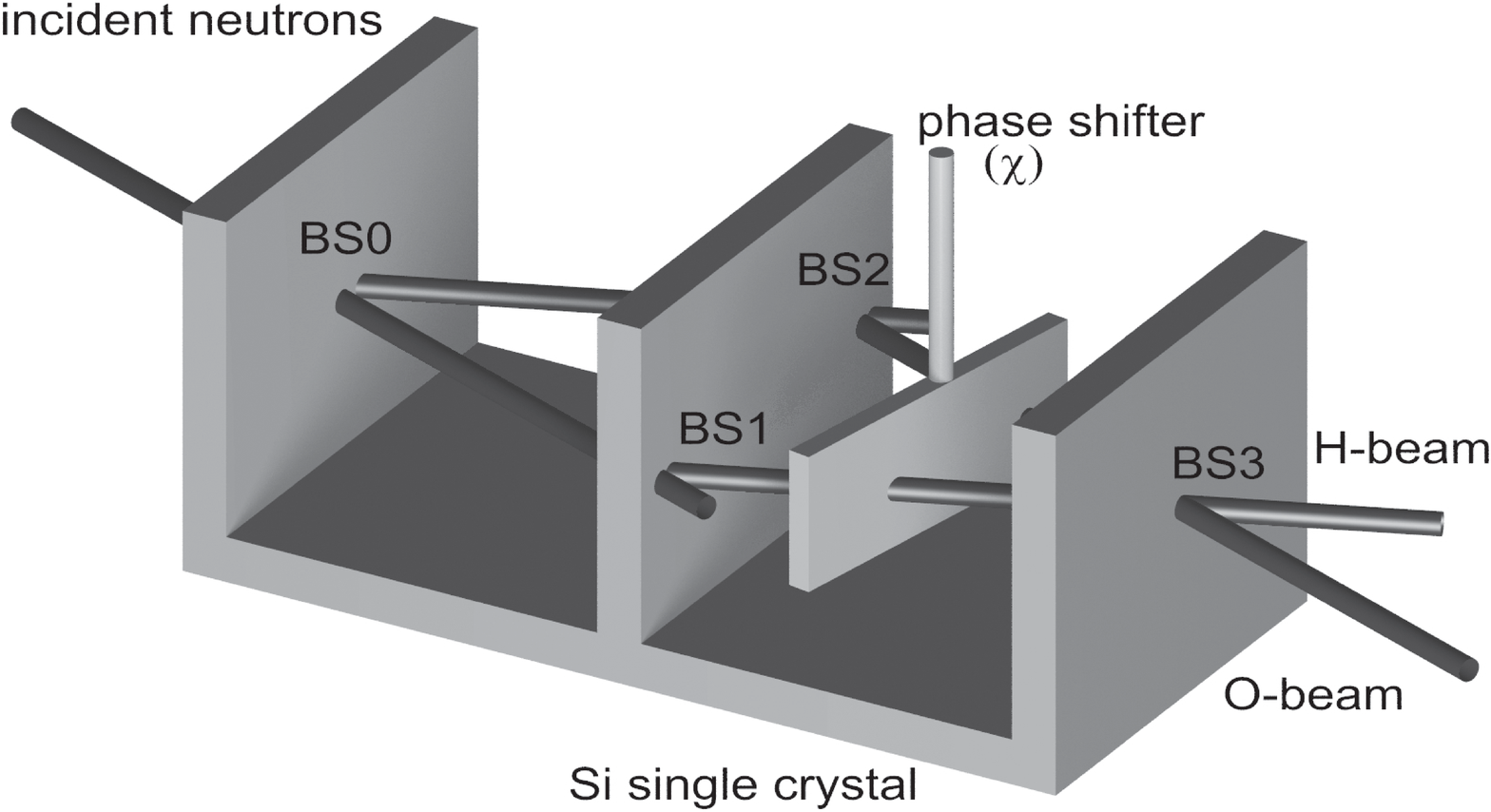}
\includegraphics[width=7cm]{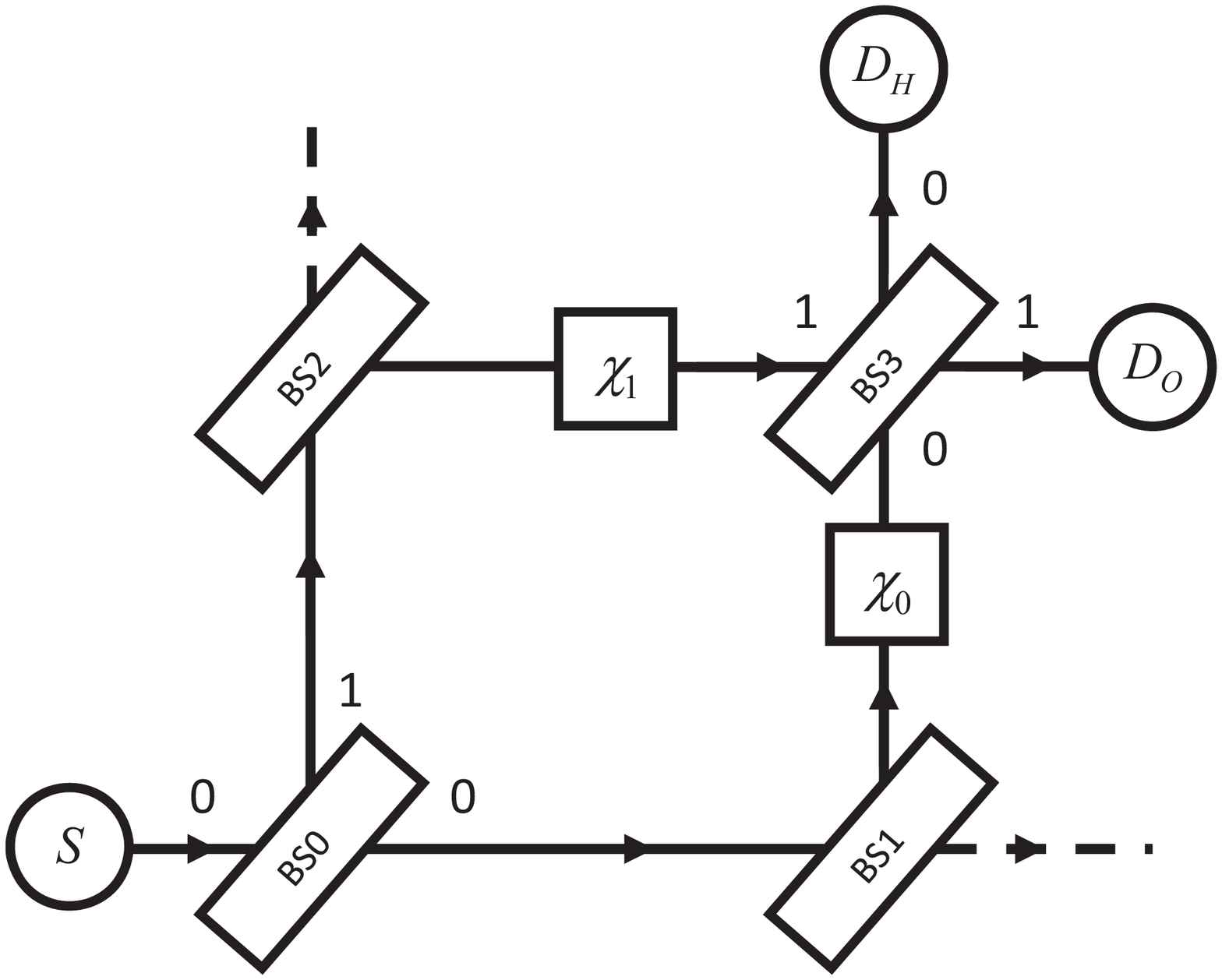}
\vspace*{8pt}
\caption{Left: Schematic picture of the silicon-perfect-crystal neutron interferometer~\cite{RAUC74a}.
BS0, $\ldots$, BS3: beam splitters; phase shifter ($\chi$): aluminum foil; neutrons transmitted by
BS1 or BS2 leave the interferometer and do not contribute to the interference
signal. Detectors count the number of neutrons in the O- and H-beam.
Right: Event-based network of the interferometer shown on the left.
S: single neutron source; BS0, $\ldots$ , BS3: beam splitters; $\chi_0$, $\chi_1$: phase shifters;
$D_O$, $D_H$: detectors counting all neutrons
leaving the interferometer via the O- and H-beam, respectively.
Neutrons transmitted by BS1 or BS2 (dashed lines) are not counted.
}
\label{fig2}
\end{center}
\end{figure}
In neutron optics there exist various realizations of the MZI type of interferometer, but here we only consider
a triple Laue diffraction type of interferometer~\cite{RAUC74a,RAUC00,HASE11}.
Figure~\ref{fig2} (left) shows the experimental configuration~\cite{RAUC74a}. The three parallel crystal plates act as beam splitters
and are assumed to be identical, which means that they have the same transmission and
reflection properties~\cite{RAUC00}. The three plates have to be parallel to high
accuracy~\cite{RAUC74a} and the whole device needs to be protected from vibrations in order to observe
interference~\cite{KROU00}.
The rotatable-plate phase shifter between the second and third plate can for example be an aluminum foil which hardly absorbs neutrons~\cite{RAUC00}.
Minute rotations of the foil about an axis perpendicular to the base plane of the interferometer
induce large variations in the phase difference $\chi=\chi_0-\chi_1$~\cite{RAUC00,LEMM10}.
The neutron detectors placed in the
so-called H-beam or O-beam have a very high, almost 100\%,
efficiency~\cite{RAUC00}.

A coherent monoenergetic neutron beam is split by the first plate (BS0).
Neutrons refracted by beam splitters BS1 and BS2 (second plate), playing the role of the mirrors in the optical MZI, are directed to the third plate (BS3)
thereby first passing through the phase shifter, and are detected by one of the two detectors.
Neutrons which are not refracted by BS1 and BS2 leave the interferometer and do not contribute to the detection counts.
The intensities in the O- and H-beam, obtained by counting individual neutrons
for a certain amount of time, exhibit sinusoidal variations as
a function of the phase shift $\chi$, a characteristic of interference~\cite{RAUC00}.

The experiment can be interpreted in different ways.
In the quantum-corpuscular view a wave packet is associated with each individual neutron.
At BS0 the wave packet splits in two components, one directed towards BS1 and one towards BS2.
At BS1 and BS2 these two components each split
in two. Two of the in total four components leave the interferometer
and the other two components are redirected towards each other at BS3 where they recombine. At BS3 the recombined wave
packet splits again in two components. Only one of these two components triggers
a detector. It is a mystery how four components of a wave packet can
conspire to do such things. Assuming that only a neutron, not
merely a part of it can trigger the nuclear reaction that causes
the detector to ``click'', on elementary logical grounds, the argument
that was just given rules out a wave-packet picture for
the individual neutron (invoking the wave function collapse only
adds to the mystery).

In the statistical interpretation of quantum mechanics there is no such conflict of interpretation~\cite{BALL03,NIEU13}.
As long as we consider
descriptions of the statistics of the experiment with many
neutrons, we may think of one single ``probability wave''
propagating through the interferometer and as the statistical interpretation
of quantum theory is silent about single events, there
is no conflict with logic either~\cite{RAUC00}.
However, the statistical interpretation does not offer an explanation of the fact that each measurement gives a definite result.

Therefore, the question to answer is: Can one model single detection events that give rise to an interference pattern if the detection events are the result of non-interacting neutrons
arriving one-by-one at the detector?
In what follows we demonstrate that
it is possible to construct a logically consistent,
cause-and-effect description in terms of discrete-event, particle like
processes which produce results that agree with those
of neutron interferometry experiments (individual detection events {\bf and} an interference pattern
after many single detection events have been collected) and the quantum theory
thereof (interference pattern only).

\subsection{Event-based model}
We construct a model for the neutrons and
for the various units in the block diagram (see Fig.~\ref{fig2} (right)) representing the setup of the neutron interferometry experiment (see Fig.~\ref{fig2} (left))
\begin{itemize}
\renewcommand\labelitemi{-}
\item{{\sl Neutrons:}
A neutron is regarded as a messenger carrying a message.
In the following we use a pictorial description, which should not be taken literally, to determine in terms of macroscopic
physics the minimal content of the message.
From other neutron experiments we deduce that the neutrons have a magnetic moment and that they
move from one point in space to another within a certain time period, the time of flight~\cite{RAUC00}.
Hence, both the magnetic moment and the time of flight have to be encoded in the message.
We model the neutron as a tiny classical magnet spinning around the direction
${\mathbf m}=(\cos\delta\sin \theta, \sin\delta\sin \theta, \cos \theta)$,
relative to a fixed frame of reference defined by a static magnetic field.
Apart from $\delta$ and $\theta$, which completely specify the magnetic moment of the netron, a third degree of freedom is required to
specify the time of flight.
This is conveniently done by representing the message by the two-dimensional unit vector
\begin{equation}
{\mathbf u}=(e^{i\psi^{(1)}}\cos (\theta/2), e^{i\psi^{(2)}}\sin (\theta/2)),
\label{neutron}
\end{equation}
where
$\psi^{(i)} =\nu t +\delta_i$, for $i=1,2$ and $\delta=\delta_1-\delta_2=\psi^{(1)}-\psi^{(2)}$.
Here, $t$ specifies the time of flight of the neutron
and $\nu$ is an angular frequency which is characteristic for a neutron
that moves with a fixed velocity $v$.
The direction of the magnetic moment ${\mathbf m}$ follows from Eq.~(\ref{neutron}) through ${\mathbf m}={\mathbf u}^T {\mathbf \sigma} {\mathbf u}$,
where $\mathbf {\sigma}= (\sigma^x,\sigma^y,\sigma^z)$ with $\sigma^x, \sigma^y, \sigma^z$ denoting the Pauli matrices
(here we use the isomorphism between the algebra of Pauli matrices and rotations in three-dimensional space).

Within the present model, the state of the neutron is fully determined
by the angles $\psi^{(1)}$, $\psi^{(2)}$ and $\theta$ and by rules (to be specified), by which these angles change
as the neutron travels through the network of devices.

The rules are as follows.
A messenger with message ${\mathbf u}$ at time $t$ and position ${\mathbf r}$
that travels with velocity $v$, along the direction ${\mathbf q}$ during a time interval
$t^{\prime} - t$, changes its message according to $\psi^{(i)}\leftarrow \psi^{(i)}+\phi$ for $i=1,2$, where
$\phi =\nu(t^{\prime}-t)$.
Hence, the new message becomes ${\mathbf w}=e^{i\phi} {\mathbf u}$.
In the presence
of a magnetic field ${\mathbf B}=(B_x,B_y,B_z)$, the magnetic moment
rotates about the direction of ${\mathbf B}$ according to the
classical equation of motion $d{\mathbf m}/dt ={\mathbf m}\times {\mathbf B}$. Hence, in a magnetic field the message ${\mathbf u}$ is changed into the message
${\mathbf w}=e^{ig\mu_NT\mathbf {\sigma}\cdot {\mathbf B}/2} {\mathbf u}$, where $g$ denotes the neutron $g$-factor, $\mu_N$ the nuclear magneton and $T$
the time during which the neutron experiences the magnetic field.
}
\item{{\sl Source:}
From the experiment we deduce that it is very unlikely that there is more than one neutron in the interferometer at a time~\cite{RAUC00}.
Therefore, we model the source such that it creates messengers one-by-one.
The source waits until the messenger's message has been processed by the detector before creating the next messenger.
Hence, there can be no direct communication between the messengers.
When the source creates a messenger, its message is initialized.
This means that the three angles $\psi^{(1)}$, $\psi^{(2)}$ and $\theta$ are specified.
The specification depends on the type of source that has to be simulated.
A monochromatic
beam of incident neutrons is assumed to consist of neutrons that
all have the same value of $\nu$~\cite{RAUC00}.
For a fully coherent spin-polarized beam of neutrons, the three angles have different but the same values for all the messengers being created.
Hence, three random numbers are used to specify $\psi^{(1)}$, $\psi^{(2)}$ and $\theta$ for all messengers.
}
\item{{\sl Beam splitters} BS0, $\ldots $ , BS3{\sl :}
To model the beam splitter for a neutron beam we
exploit the similarity between the magnetic
moment of the neutron and the polarization of a photon.
Therefore we use a model for the beam splitter that is similar to the one used
for polarized photons, thereby assuming that neutrons with spin up and
spin down have the same reflection and transmission
properties, while photons with horizontal and vertical
polarization have different reflection and transmission
properties~\cite{BORN64}. The coordinate system is defined by the static magnetic guiding field.

In optics, dielectric plate beam splitters are often used to partially transmit and partially reflect an incident light beam.
From classical electrodynamics we know that if an electric field is applied to a dielectric material the
material becomes polarized~\cite{BORN64}.
Assuming a linear response, the polarization vector of the material is given by
${\mathbf P}(\omega)={\mathbf \chi}(\omega){\mathbf E}(\omega)$ for a monochromatic wave with
frequency $\omega$. In the time domain, this relation expresses the fact that the material
retains some memory about the incident field, ${\mathbf \chi}(\omega)$ representing the memory kernel
that is characteristic for the material used.
We use a similar kind of memory effect in our algorithm to model the beam splitter for neutrons.
Note that memory effects are also essential for uncertainty and quantization phenomena observed in a system of
droplets bouncing on a liquid bath~\cite{COUD10,COUD11,COUD12}.

A beam splitter has two input and two output channels labeled by 0 and 1 (see Fig~\ref{fig2} (right)).
Note that in the neutron interferometry experiment, for beamsplitter BS0 only entrance port $k=0$ is used.
In the event-based model, the beam splitter has two internal registers ${\mathbf R}_{k,n}=(R_{0,k,n},R_{1,k,n})$ (one for each input channel)
with $R_{i,k,n}$ for $i=0,1$ representing a complex number, and an
internal vector ${\mathbf v}_n=(v_{0,n},v_{1,n})$ with the additional constraints that $v_{i,n}\ge0$ for $i=0,1$ and that $v_{0,n}+v_{1,n}=1$.
As we only have two input channels, the latter constraint can be used
to recover $v_{1,n}$ from the value of $v_{0,n}$.
These three two-dimensional vectors ${\mathbf v}_n$, ${\mathbf R}_{0,n}$ and ${\mathbf R}_{1,n}$ are labeled by the message number
$n$ because their content is updated every time the beam splitter receives a message.
Before the simulation starts we set ${\mathbf v}_0 = (v_{0,0}, v_{1,0}) = (r, 1 - r)$, where $r$ is a uniform pseudorandom
number. In a similar way we use pseudo-random numbers to initialize ${\mathbf R}_{0,0}$ and ${\mathbf R}_{1,0}$.

When the $n$th messenger carrying the message $\mathbf{u}_{k,n}$ arrives at entrance port $k=0$ or $k=1$ of the beam splitter,
the beam splitter first stores the message in the corresponding register ${\mathbf R}_{k,n}$ and updates its internal vector according to the rule
\begin{equation}
{\mathbf v}_n=\gamma {\mathbf v}_{n-1}+(1-\gamma){\mathbf q}_n,
\label{internalBS}
\end{equation}
where $0<\gamma<1$ and ${\mathbf q}_n=(1,0)$ (${\mathbf q}_n=(0,1)$) if the $n$th event occurred on channel $k=0$ ($k=1$).
Note that exactly ten floating point numbers can be stored in the registers ${\mathbf R}_{k,n}$ and the internal vector ${\mathbf v}_n$.
The beam splitter stores information about the last message only. The information
carried by earlier messages is overwritten by updating the
internal registers and internal vector.
By construction $v_{i,n}\ge 0$ for $i=0,1$ and $v_{0,n}+v_{1,n} =1$. Hence the update rule Eq.~(\ref{internalBS})
preserves the constraints on the internal vector. Obviously,
these constraints are necessary if we want to interpret the $v_{k,n}$
as (an estimate of) the relative frequency for the occurrence of an event of type $k$.

From Eq.~(\ref{internalBS}), one could
say that the internal vector ${\mathbf v}$ (corresponding to the material polarization ${\mathbf P}$ in classical electromagnetism) is the response of the beam splitter to the
incoming messages represented by the vectors ${\mathbf q}$ (corresponding to the electric field ${\mathbf E}$).
Elsewhere~\cite{JIN10b} we have shown that the update rule Eq.~(\ref{internalBS}) for the internal vector leads to the Debye model for the interaction
between material and electric field.
In other words, the beam splitter ``learns'' so to speak from the information
carried by the messengers. The characteristics of the learning
process depend on the parameter $\gamma$ (corresponding to the
response function $\chi$).
Since this learning mechanism is completely deterministic we call this type of message processor a deterministic learning machine (DLM)
with a parameter $\gamma$ that controls the pace and accuracy of the learning process.

The next step is to use the ten numbers stored in ${\mathbf R}_{k,n}$ and ${\mathbf v}_n$ to calculate four complex numbers
\begin{eqnarray}
\left(
\begin{array}{c}
h_{0,n}\\
h_{1,n}\\
h_{2,n}\\
h_{3,n}
\end{array}
\right)
&=&
\left(
\begin{array}{cccc}
\sqrt{{\cal T}}&i\sqrt{{\cal R}}&0&0\\
i\sqrt{{\cal R}}&\sqrt{{\cal T}}&0&0\\
0&0&\sqrt{{\cal T}}&i\sqrt{{\cal R}}\\
0&0&i\sqrt{{\cal R}}&\sqrt{{\cal T}}
\end{array}
\right)\nonumber \\
&&\times
\left(
\begin{array}{cccc}
\sqrt{v_{0,n}}&0&0&0\\
0&\sqrt{v_{1,n}}&0&0\\
0&0&\sqrt{v_{0,n}}&0\\
0&0&0&\sqrt{v_{1,n}}
\end{array}
\right)
\left(
\begin{array}{c}
R_{0,0,n}\\
R_{0,1,n}\\
R_{1,0,n}\\
R_{1,1,n}
\end{array}
\right)\nonumber \\
&=&
\left(
\begin{array}{c}
\sqrt{v_{0,n}}\sqrt{{\cal T}}R_{0,0,n}+i\sqrt{v_{1,n}}\sqrt{{\cal R}}R_{0,1,n}\\
i\sqrt{v_{0,n}}\sqrt{{\cal R}}R_{0,0,n}+\sqrt{v_{1,n}}\sqrt{{\cal T}}R_{0,1,n}\\
\sqrt{v_{0,n}}\sqrt{{\cal T}}R_{1,0,n}+i\sqrt{v_{1,n}}\sqrt{{\cal R}}R_{1,1,n}\\
i\sqrt{v_{0,n}}\sqrt{{\cal R}}R_{1,0,n}+\sqrt{v_{0,n}}\sqrt{{\cal T}}R_{1,1,n}
\end{array}
\right)
,
\label{PBS1neutron}
\end{eqnarray}
where the reflectivity $\cal{R}$ and transmissivity ${\cal T}=1-\cal{R}$ are real numbers
which are considered to be parameters to be determined from experiment.

It is clear that the computation of the numbers $h_{0,n},\ldots,h_{3,n}$ plays the role
of the matrix-vector multiplication in the quantum-theoretical description of the beam splitter.
Note however that the input and output amplitudes are constructed event-by-event and only under certain conditions
($\gamma\rightarrow 1^-$, sufficiently large number of input events $N$, stationary sequence of input events) they correspond
to their quantum theoretical counterparts.

In the final step $h_{0,n},\ldots , h_{3,n}$ are used to create an output event.
To this end a uniform random number $0<r_n<1$ is generated.
If $|h_{0,n}|^2+|h_{2,n}|^2 > r_n$, a message
\begin{equation}
{\mathbf w}_{0,n}=(h_{0,n},h_{2,n})/\sqrt{|h_{0,n}|^2+|h_{2,n}|^2},
\end{equation}
is sent through output channel 1. Otherwise
a message
\begin{equation}
{\mathbf w}_{1,n}=(h_{1,n},h_{3,n})/\sqrt{|h_{1,n}|^2+|h_{3,n}|^2},
\end{equation}
is sent through output channel 0.
}
\item{{\sl Phase shifter $\chi_0$, $\chi_1$:}
A phase shifter is simulated without DLM.
The device has only one input and one output port and
transforms the $n$th input message ${\mathbf u}_n$ into an output message
\begin {equation}
{\mathbf w}_n=e^{i\chi_j}{\mathbf u}_n\quad j=0,1,
\end{equation}
}
where $\chi_j$ denotes the phase.
\item{{\sl Detector:}
Detectors count all incoming neutrons and hence have a detection efficiency of 100\%.
This is an idealization of real neutron detectors which can have a detection efficiency of 99\% and more~\cite{KROU00}.
After detection the neutron is destroyed.
}
\end{itemize}

\subsection{Simulation results}
\begin{figure}[pt]
\begin{center}
\includegraphics[width=7.5cm]{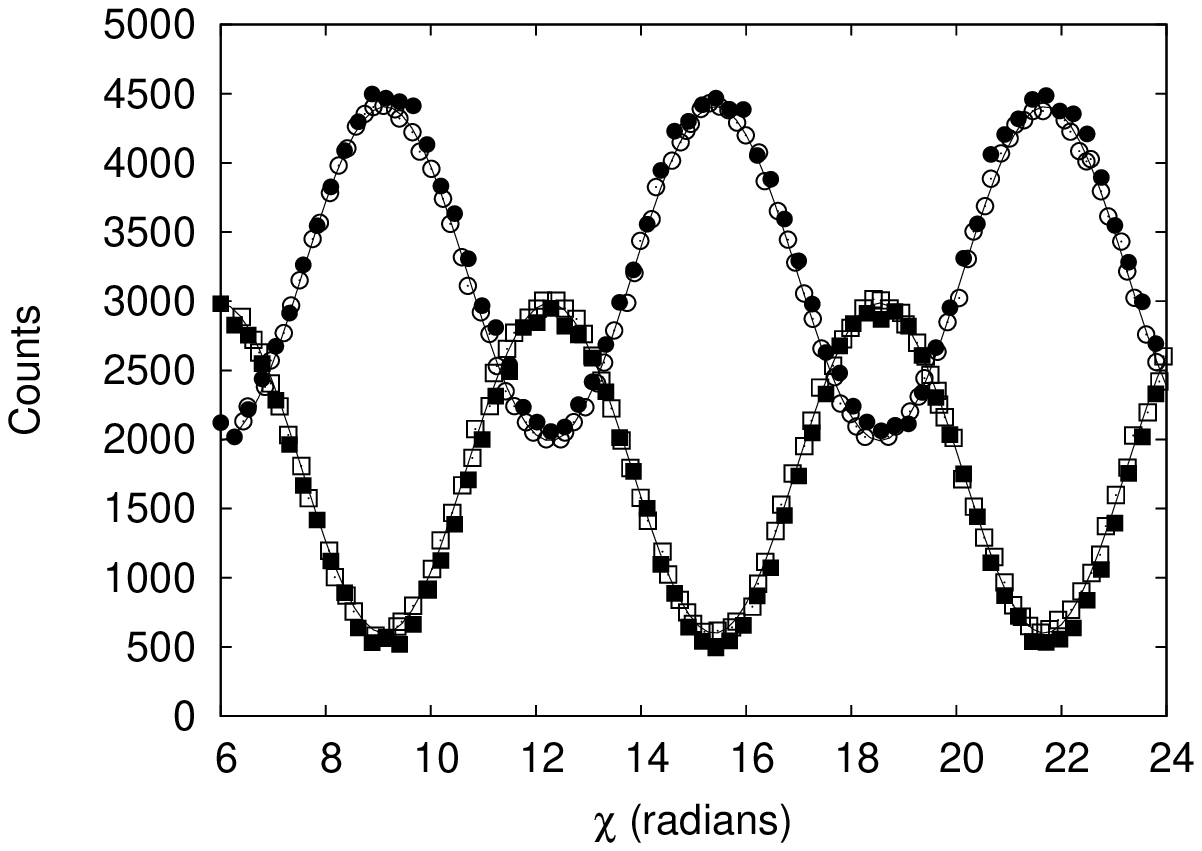}
\includegraphics[width=7.5cm]{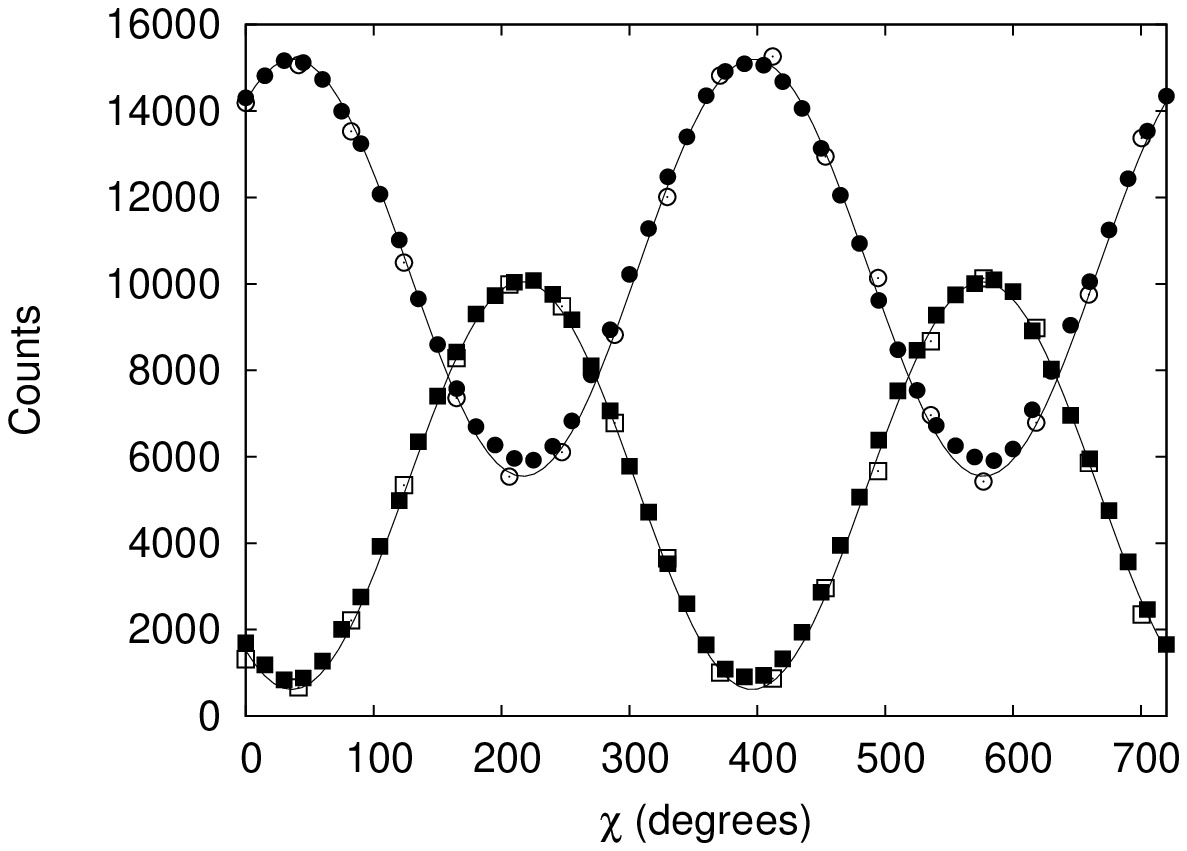}
\vspace*{8pt}
\caption{%
Comparison between the counts of neutrons per second and per square cm in the beams of a neutron interferometry experiment
(open symbols) and the number of neutrons per sample leaving the interferometer in an event-by-event simulation
(solid symbols). Circles: counts in the H-beam; squares: counts in the O-beam.
Lines through the data points are guides to the eye.
Left: experimental data extracted from
Fig.~2 of Ref.~\cite{KROU00}; simulation parameters: ${\cal R} = 0.22$, $\gamma= 0.5$, number of incident neutrons per angle $\chi$:
$N = 22727$.
Right: experimental data provided by H. Lemmel and H. Rauch (private communication);
simulation parameters: ${\cal R} = 0.22$, $\gamma= 0.7$, number of incident neutrons per angle $\chi$: $N = 72727$.
The parameters ${\cal R}$ and $\gamma$ have been adjusted by hand to obtain a good fit.
}
\label{fig3}
\end{center}
\end{figure}

If $\gamma$, the parameter which controls the learning pace of a DLM-based processor, approaches one the event-by-event simulation
reproduces the results of quantum theory ~\cite{RAED05b,RAED05d,MICH11a,RAED12b}.
In the neutron interferometry experiment the parameter $\gamma$, controlling the learning pace of the beam splitters BS0, $\ldots$, BS3,
can be used to account for imperfections of the neutron interferometer~\cite{RAED12b}.

In Fig.~\ref{fig3}, we present a comparison of simulation data with experimental data using two different
experimental data sets.
For Fig.~\ref{fig3} (left) we use experimental data of an experiment performed in 2000 by Kroupa {\sl et al.}~\cite{KROU00}.
For Fig.~\ref{fig3} (right) we use data from a more recent experiment.
Both figures show that the event-based simulation model reproduces, quantitatively, the experimental results.
We only had to use different values for $\gamma$ to get a good fit to the experimental curves.
For the oldest experiment we used $\gamma =0.5$ and for the more recent one $\gamma =0.7$.
This could indicate that the more recent experimental setup suffers from less imperfections (note also the difference in neutron counts) so that a better agreement with quantum theory is found.
In this respect $\gamma$ could also be interpreted as a kind of ``(de)coherence parameter''.

In conclusion, the event-based simulation approach models the experimentally observed one-by-one build-up process of the interference pattern
as a chronological sequence of different events including the source emitting neutrons, the neutrons interacting with the beam splitters and
the neutrons arriving at the detector thereby generating detection events.

\section{Neutron spin-path entanglement}
Now that we have given an event-based description of the fact that individual neutrons which do not interact with each other exhibit the collective behavior
that gives rise to an interference pattern, another intriguing
question to answer is whether one can model single detection events that give rise to entanglement, another quantum mechanical phenomenon, if the detection events are the result of
non-interacting neutrons arriving one-by-one at the detector.

The single-neutron interferometry experiment of Hasegawa {\it et al.}~\cite{HASE03}
demonstrates that the correlation between
the spatial and spin degree of freedom of neutrons violates a Bell-CHSH (Clauser-Horne-Shimony-Holt) inequality.
In this section we construct an event-based model that reproduces this correlation.
We show that the event-based model reproduces
the exact results of quantum theory if $\gamma\rightarrow 1^-$ and that by changing $\gamma$ it can also reproduce the numerical
values of the correlations, as measured in experiments~\cite{HASE03,BART09}.
Note that this Bell-test experiment involves two degrees of freedom of one particle, while the Einstein-Podolsky-Rosen-Bohm (EPRB) thought experiment~\cite{BOHM51}
and EPRB experiments with single photons~\cite{WEIH98,WEIH00,HNIL02,AGUE09} involve two degrees of freedom of two particles.
Hence, the Bell-test experiment with single neutrons is not performed according to the CHSH protocol~\cite{CLAU69} because the two degrees of freedom
of one particle are not manipulated and measured independently.

\begin{figure}[t]
\begin{center}
\includegraphics[width=8cm]{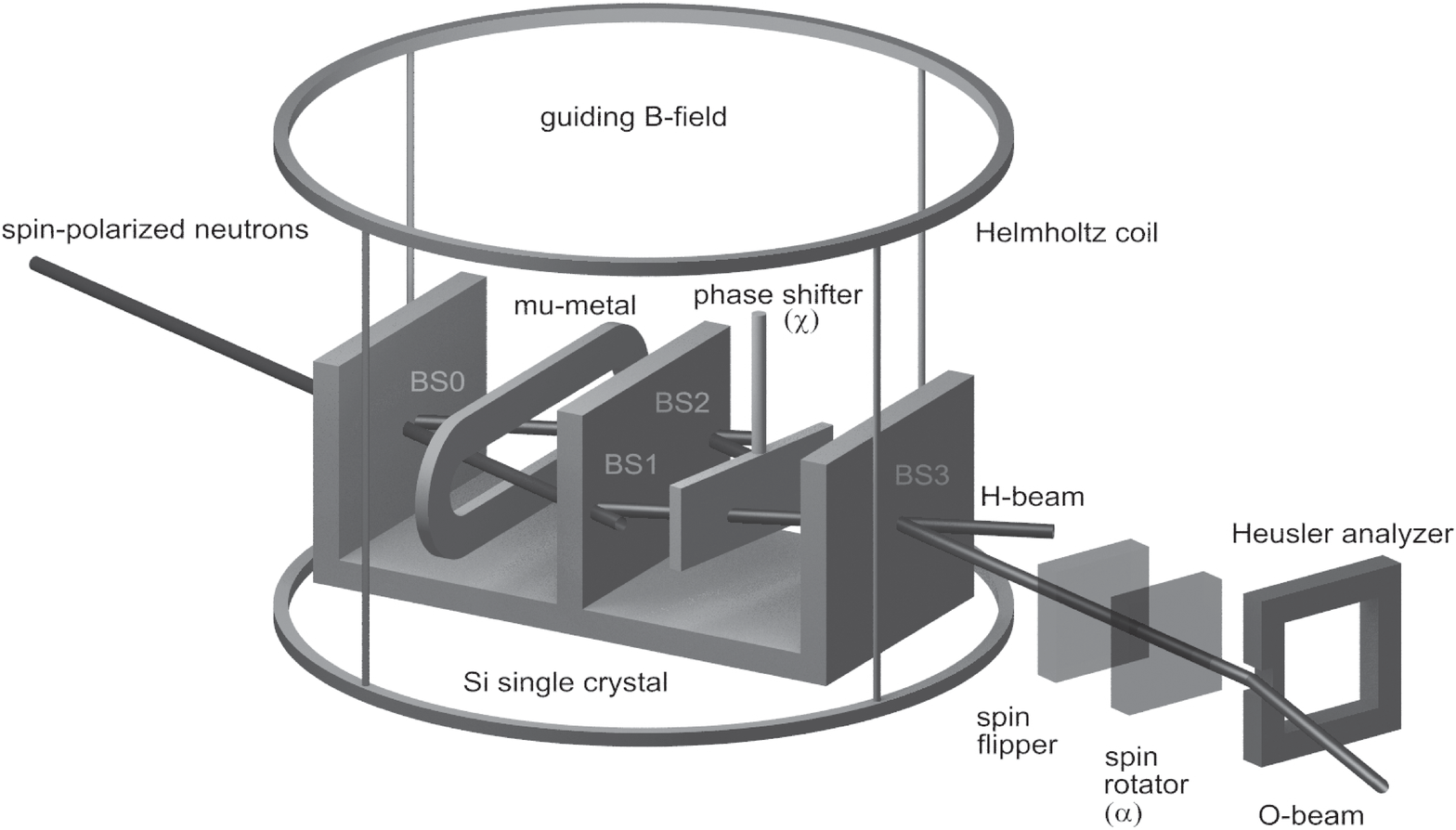}
\includegraphics[width=7cm]{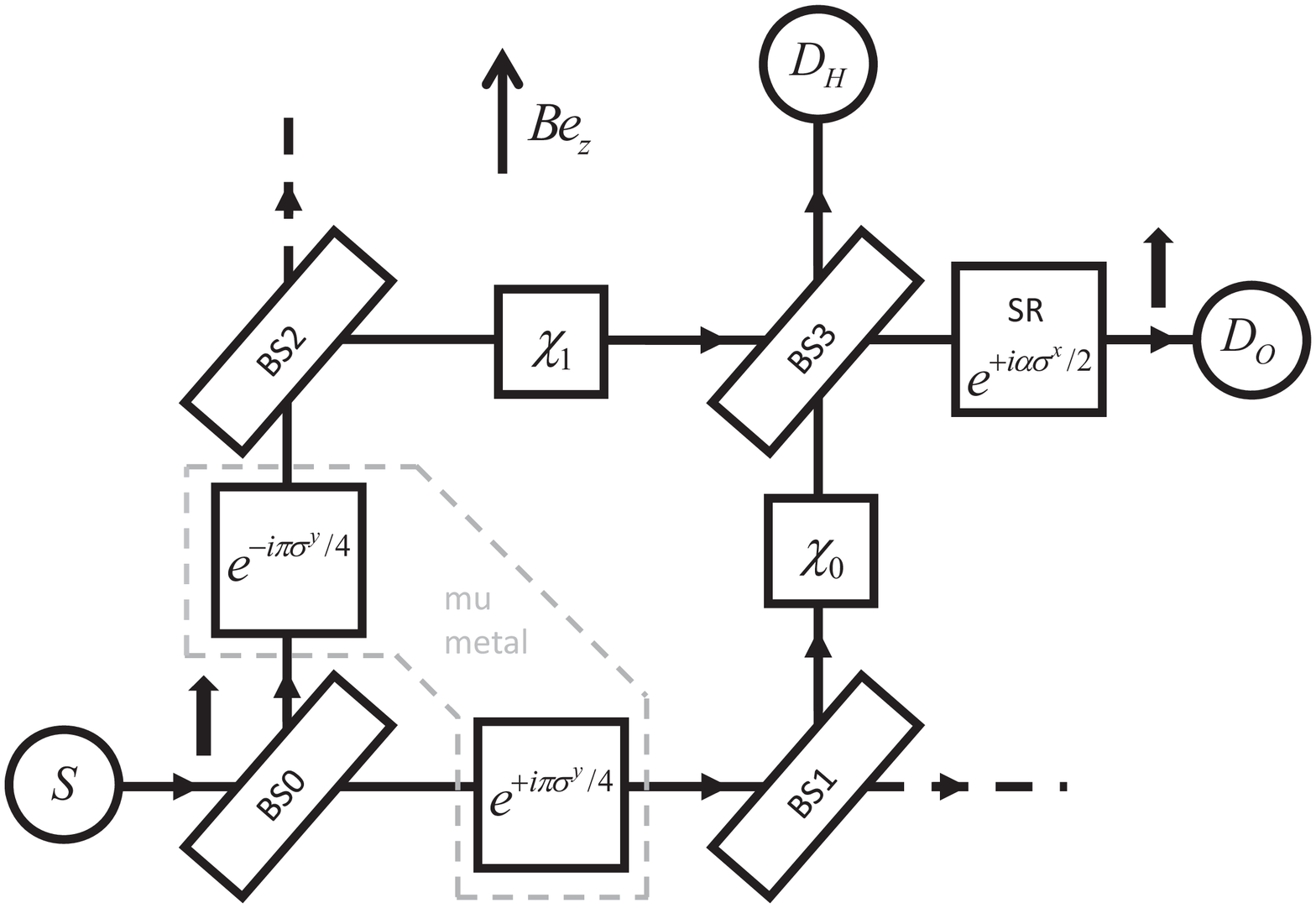}
\caption{%
Left: Schematic picture of the single-neutron interferometry experiment to test a Bell inequality violation (see also Fig.~1 in Ref.~\cite{HASE03}).
BS0, $\ldots$, BS3: beam splitters; phase shifter $\chi$: aluminum foil; neutrons transmitted by BS1 or BS2 leave the
interferometer and do not contribute to the interference signal. Detectors count the number of neutrons in the
O- and H-beam.
Right: Event-based network of the experimental setup shown on the left.
$B{\mathbf e}_z$: magnetic guiding field along the $z$-direction, S: single neutron source; BS0, $\ldots$ , BS3: beam splitters;
$e^{+i\pi\sigma^y/4}$, $e^{-i\pi\sigma^y/4}$: spin rotators modeling the action of a mu metal;
$\chi_0$, $\chi_1$: phase shifters;
SR $e^{i\alpha\sigma^x/2}$: spin rotator;
$D_O$, $D_H$: detectors counting all neutrons leaving
the interferometer via the O- and H-beam, respectively.
Neutrons leaving the interferometer via
the dashed lines are not counted.
}
\label{fig4}
\end{center}
\end{figure}

Figure~\ref{fig4} (left) shows a schematic picture of the single-neutron interferometry experiment.
Incident neutrons pass through a magnetic-prism polarizer (not shown) which produces two spatially separated beams of
neutrons with their magnetic moments aligned parallel (spin up), respectively anti-parallel (spin down) with respect
to the magnetic axis of the polarizer which is parallel to the guiding field ${\mathbf B}$. The spin-up neutrons
impinge on a silicon-perfect-crystal interferometer~\cite{RAUC00}. On leaving the first beam splitter BS0,
neutrons are transmitted or refracted.
A mu-metal spin-turner changes the orientation of the magnetic moment of the neutron from parallel to perpendicular to the guiding field ${\mathbf B}$.
Hence, the magnetic moment of the neutrons following path H (O) is rotated by $\pi/2$ ($-\pi/2$)
about the $y$ axis. Before the two paths join at the entrance plane of beam splitter BS3, a difference between the time of flights
along the two paths can be manipulated by a phase shifter $\chi$. The neutrons
which experience two refraction events when passing through the interferometer form the O-beam and are analyzed by sending them through
a spin rotator and a Heusler spin analyzer. If necessary, to induce an extra spin rotation of $\pi$, a spin flipper is placed between
the interferometer and the spin rotator. The neutrons that are selected by the Heusler spin analyzer are counted with a
neutron detector (not shown) that has a very high efficiency ($\approx 99\%$).
Note that neutrons which are not refracted by the second plate leave the interferometer
without being detected.

The single-neutron interferometry experiment yields the count rate $N(\alpha,\chi)$ for the spin-rotation angle $\alpha$ and
the difference $\chi$ of the phase shifts of the two different paths in the interferometer~\cite{HASE03}.
The correlation $E(\alpha,\chi)$ is defined by~\cite{HASE03}

\begin{equation}
E(\alpha,\chi)=\frac{N(\alpha,\chi)+N(\alpha+\pi,\chi+\pi)-N(\alpha+\pi,\chi)-N(\alpha,\chi+\pi)}
{N(\alpha,\chi)+N(\alpha+\pi,\chi+\pi)+N(\alpha+\pi,\chi)+N(\alpha,\chi+\pi)}.
\end{equation}

\subsection{Event-based model}
A minimal, discrete event simulation model of the single-neutron interferometry experiment requires a specification of the
information carried by the neutrons, of the algorithm that simulates the source and the interferometer components
(see Fig.~\ref{fig4} (right)), and of the procedure to analyze the data.
Various ingredients of the simulation model have already been described in Sect.~2.
In the following, we specify the action of the remaining components.

\begin{itemize}
\renewcommand\labelitemi{-}
\item{{\sl Magnetic-prism polarizer:} transforms the
unknown magnetic moment of an incoming neutron into
a magnetic moment that is either parallel (spin up) or antiparallel
(spin down) with respect to the $z$-axis (which by definition
is parallel to the guiding field ${\mathbf B}$). In the experiment, only a
neutron with spin up is injected into the interferometer. Therefore,
as a matter of simplification, we assume in the event-based simulation that the source S only creates messengers with spin up.
Hence, we assume that $\theta =0$ in Eq.~(\ref{neutron}).
}
\item{{\sl Mu metal spin turner:}
rotates the magnetic moment of a neutron that follows the H-beam (O-beam) by $\pi/2$ ($-\pi/2$) about the $y$ axis.
In the event-based simulation, the processor (not a DLM) that accomplishes this
takes as input the direction of the magnetic moment,
represented by the message $\mathbf u$ and performs
the rotation ${\mathbf u}\leftarrow e^{i\pm\pi\sigma ^y/4}{\mathbf u}$.
We emphasize that we use Pauli matrices as a convenient tool
to express rotations in three-dimensional space, not because in quantum theory
the magnetic moment of the neutron is represented by spin-1/2
operators.
}
\item{{\sl Spin-rotator and spin-flipper:}
rotates the magnetic moment of a neutron by an angle $\alpha$ about the $x$ axis.
We model this component by a processor (not a DLM) that takes as input the message $\mathbf u$ and performs the rotation
${\mathbf u}\leftarrow e^{+i\alpha\sigma^x/2}{\mathbf u}$.
The spin flipper is a spin rotator with $\alpha=\pi$.
}
\item{{\sl Heusler spin analyzer:}
In the experiment the spin analyzer selects neutrons with spin up, after which they are counted by a detector.
The model of this component is a device that lets the neutron pass if $r\le (1+m_z)/2$, where $0<r<1$ is a uniform pseudo random number, and destroys the neutron otherwise.
}
\end{itemize}
\begin{figure}[t]
\begin{center}
\includegraphics[width=7.5cm]{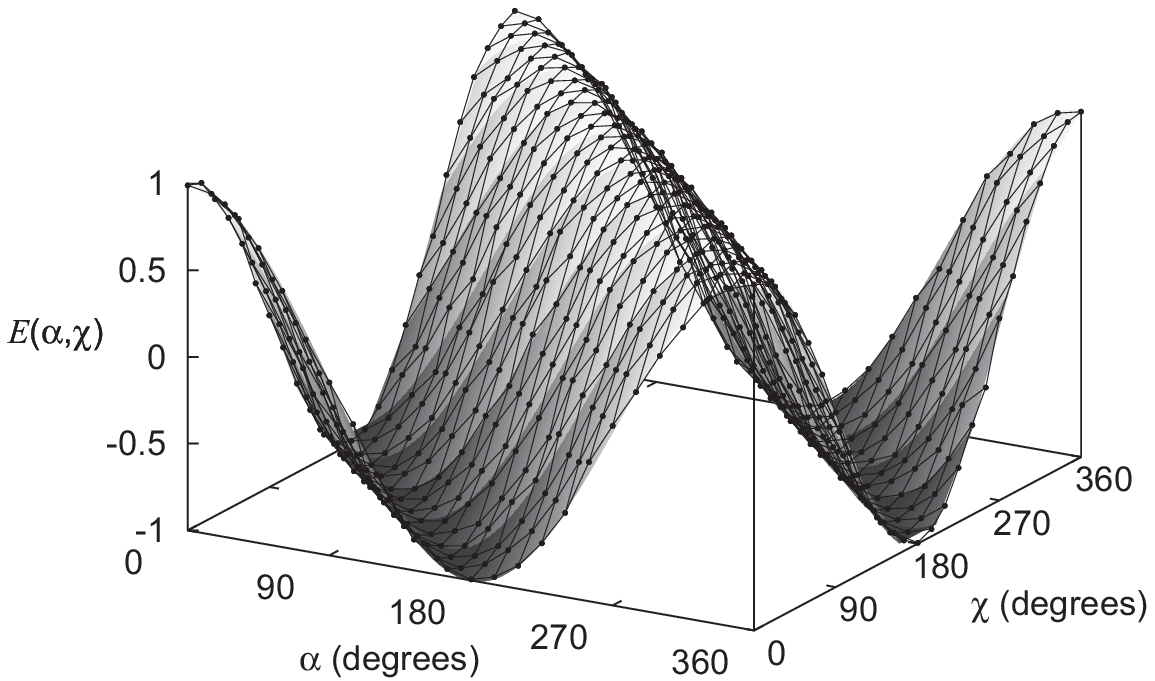}
\includegraphics[width=7.5cm]{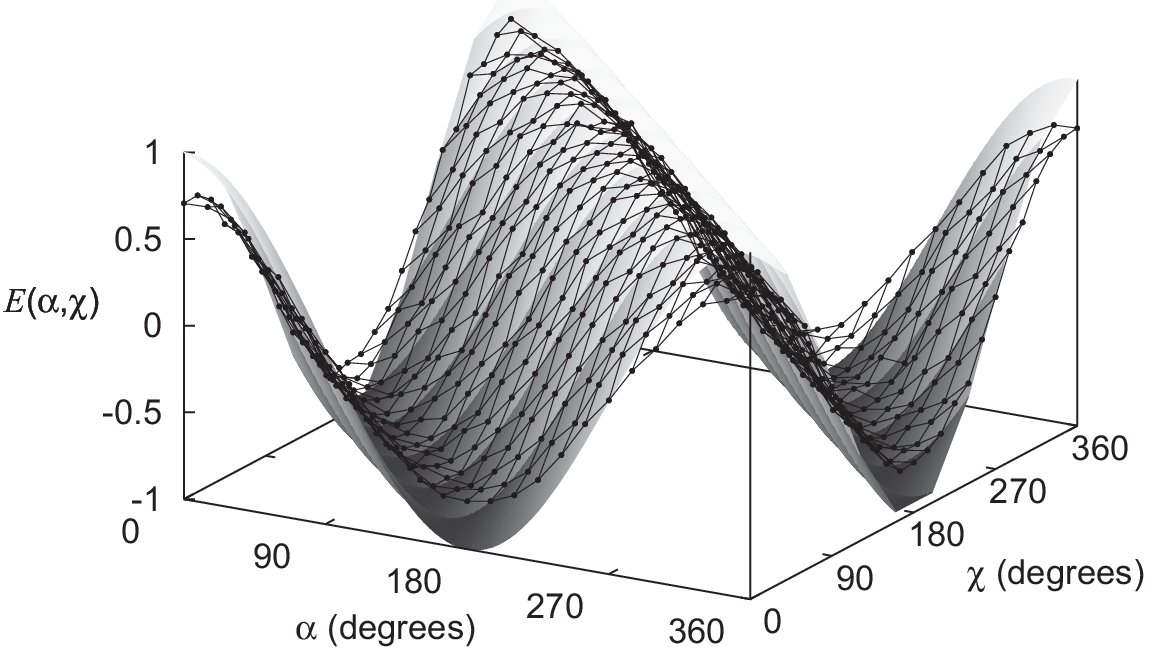}
\caption{%
Correlation $E(\alpha,\chi)$ between spin and path degree of freedom as obtained from an event-based simulation of the experiment
depicted in Fig.~\ref{fig3}. Solid surface: $E(\alpha,\chi)=\cos(\alpha+\chi)$ predicted by quantum theory; dots: simulation data.
Model parameters: ${\cal R}=0.2$, $\gamma=0.99$ (left) and $\gamma =0.55$ (right).
The lines connecting the markers are guides to the eye.
For each pair $(\alpha,\chi)$, four times 10000 particles were used to determine the four counts $N(\alpha,\chi)$, $N(\alpha+\pi,\chi+\pi)$,
$N(\alpha,\chi+\pi)$ and $N(\alpha+\pi,\chi+\pi)$.
}
\label{fig5}
\end{center}
\end{figure}

\subsection{Simulation results}
In Fig.~\ref{fig5} (left) we present simulation results for the correlation $E(\alpha,\chi)$, assuming that the experimental conditions
are very close to ideal and compare them to the quantum theoretical result~\cite{HASE03}
$E_\mathrm{O}(\alpha,\chi)=\cos(\alpha+\chi)$ with $\chi=\chi_0-\chi_1$,
independent of the reflectivity ${\cal R}$ of the beam splitters (which have been assumed to be identical).
The fact that $E_\mathrm{O}(\alpha,\chi)=\cos(\alpha+\chi)$ implies that the state of the neutron
cannot be written as a product of the state of the spin and the phase.
In other words, in quantum language, the spin- and phase-degree-of-freedom are entangled~\cite{BASU01,HASE03}.
If the mu-metal would rotate the spin about the $x$-axis instead of about the $y$-axis,
we would find $E_\mathrm{O}(\alpha,\chi)=\cos\alpha\cos\chi$, a typical expression for a quantum system in a product state.

As shown by the markers in Fig.~\ref{fig5} (left), disregarding the small statistical fluctuations,
there is close-to-perfect agreement
between the event-based simulation data for nearly ideal experimental conditions ($\gamma =0.99$ and ${\cal R}=0.2$) and quantum theory.
However, the laboratory experiment suffers from unavoidable imperfections, leading to a reduction and distortion of the interference fringes~\cite{HASE03}.
In the event-based approach it is trivial to incorporate mechanisms for different sources of imperfections by modifying or adding update rules.
However, to reproduce the available data it is sufficient to use the parameter $\gamma$ to control the deviation from the quantum theoretical result.
For instance, for $\gamma=0.55$, ${\cal R}=0.2$ the simulation results for $E(\alpha , \chi )$ are shown in Fig.~\ref{fig5} (right).

In order to quantify the difference between the simulation results, the experimental results and quantum theory
it is customary to form the Bell-CHSH function~\cite{BELL93,CLAU69}
\begin{equation}
S=S(\alpha,\chi,\alpha^{\prime},\chi^{\prime})
= E_\mathrm{O}(\alpha,\chi)
+E_\mathrm{O}(\alpha,\chi^{\prime})
-E_\mathrm{O}(\alpha^{\prime},\chi)
+E_\mathrm{O}(\alpha^{\prime},\chi^{\prime})
,
\label{app8}
\end{equation}
for some set of experimental settings $\alpha$, $\chi$, $\alpha^{\prime}$, and $\chi^{\prime}$.
If the quantum system can be described by a product state, then $|S|\le2$.
If $\alpha=0$, $\chi=\pi/4$, $\alpha^{\prime}=\pi/2$, and $\chi^{\prime}=\pi/4$, then
$S\equiv S_{max}=2\sqrt{2}$, the maximum value allowed by quantum theory~\cite{CIRE80}.

For $\gamma =0.55$, ${\cal R}=0.2$ the simulation results yield
$S_{max}=2.05$, in excellent agreement with the value $2.052\pm 0.010$
obtained in experiment~\cite{HASE03}. For $\gamma=0.67$, ${\cal R}=0.2$ the simulation yields $S_{max}=2.30$, in excellent agreement with the value $2.291\pm 0.008$
obtained in a similar, more recent experiment~\cite{Bartosik2009}.

In conclusion, since experiment shows that $|S|>2$, according to quantum theory it is impossible
to interpret the experimental result in terms of a quantum system in the product state~\cite{BALL03}.
The system must be described by an entangled state.
However, the event-based simulation approach which makes use of classical, Einstein-local and causal event-by-event processes
can reproduce all the features of this experiment.

\section{Ozawa's error-disturbance relation}
Very recently, a neutron-optical experiment has been reported~\cite{ERHA12,SULY13} of which the results confirm a new
error-disturbance uncertainty relation~\cite{OZAW03}, derived using the theory of general quantum measurements.
In what follows we demonstrate that the event-based simulation approach can also model this single neutron experiment.

The universally valid error-disturbance uncertainty relation derived by Ozawa reads~\cite{OZAW03}

\begin{equation}
\epsilon (A)\eta (B)+\epsilon (A)\Delta (B)+\Delta (A)\eta (B) \ge |\langle\psi |[A,B]|\psi\rangle |/2,
\label{ozawa1}
\end{equation}
for every measurement and every state $|\psi\rangle$.
For a projective measurement, the error $\epsilon (A)$ on the measurement of the observable $A$ is defined as
the root mean squared of the difference
between the output operator $O_A$ actually measured and the observable $A$ to be measured and is given by
$\epsilon (A)=\|(O_A-A)|\psi\rangle\|$~\cite{ERHA12}. The disturbance $\eta (B)$ on the measurement of the observable $B$
caused by the measurement of $A$ is defined as the root mean squared of the change in
observable $B$ during the measurement and reads $\eta (B)=\|[O_A,B]|\psi\rangle\|$~\cite{ERHA12}.
The variance $\Delta^2 (X)$ of an observable $X$ is defined as $\Delta^2 (X)= \langle X^2\rangle -\langle X\rangle^2$.
Uncertainty relation Eq.~(\ref{ozawa1}) is a ``universally valid'' version of the inequality
\begin{equation}
\epsilon (A)\eta (B) \ge |\langle\psi |[A,B]|\psi\rangle |/2,
\label{heisenberg}
\end{equation}
which is the result of an interpretation~\cite{OZAW03,ERHA12,SULY13,FUJI12,FUJI13a}
of Heisenberg 's original writings~\cite{HEIS27},
whereby it is assumed, without solid justification, that $\epsilon (A)$ and $\eta (B$) correspond to the
``uncertainties'' which Heisenberg had in mind, see also \cite{BUSH07,BUSH13}.
Unlike Eq.~(\ref{ozawa1}), inequality Eq.~(\ref{heisenberg}) can be violated since it lacks a mathematical
rigorous basis~\cite{BALL70}.

In the neutron experiment the idea is to first measure the error of the $x$- and then the disturbance on the
$y$-component of the neutron spin $S=1/2$ if the state $|\psi\rangle=\left|\uparrow\rangle\right.$.
Hence, the observables $A$ and $B$ in Eq.~(\ref{ozawa1}) correspond to $A=\sigma^x$ and $B=\sigma^y$, respectively.
In order to control $\epsilon (A)$ and $\eta (B)$ an error model is introduced so that instead of measuring
$A=\sigma^x$ one measures $O_A=\sigma^{\phi}=\sigma^x\cos\phi +\sigma^y\sin\phi$, where the angle $\phi$, also called the detuning angle, is
controlled by the experimenter.
The error-disturbance relation Eq.~(\ref{ozawa1}) then reads~\cite{ERHA12,SULY13}
\begin{equation}
2\sqrt{2}\cos\phi\sin (\phi /2) +2\sin (\phi /2)
+\sqrt{2}\cos\phi \ge 1.
\label{ozawa2}
\end{equation}
However, the application of Eq.(\ref{ozawa1}) to the neutron experiment is not straightforward.
With some clever manipulations~\cite{ERHA12,SULY13},
it is possible to express the unit operators that appear in Eq.(\ref{ozawa1}) in terms of dynamical
variables, the expectations of which can be extracted from the data of single-neutron experiments.
If the state of the spin-1/2 system is described by the density matrix $\rho=|z\rangle \langle z|$,
we have~\cite{ERHA12,SULY13}
\begin{eqnarray}
\epsilon^2(A)&=&2+\langle z|O_A|z\rangle +\langle -z|O_A|-z\rangle -2\langle x|O_A|x\rangle, \nonumber\\
\eta^2(B)&=& 2+\langle z|O_B|z\rangle +\langle -z|O_B|-z\rangle -2\langle y|O_B|y\rangle,
\label{ozawa3}
\end{eqnarray}
where $O_B=((1+\sigma^{\phi})\sigma^y(1+\sigma^{\phi})+(1-\sigma^{\phi})\sigma^y(1-\sigma^{\phi}))/4=
\sin\phi\sigma^{\phi}=\sin\phi O_A$.

\begin{figure}[t]
\begin{center}
\includegraphics[width=5cm]{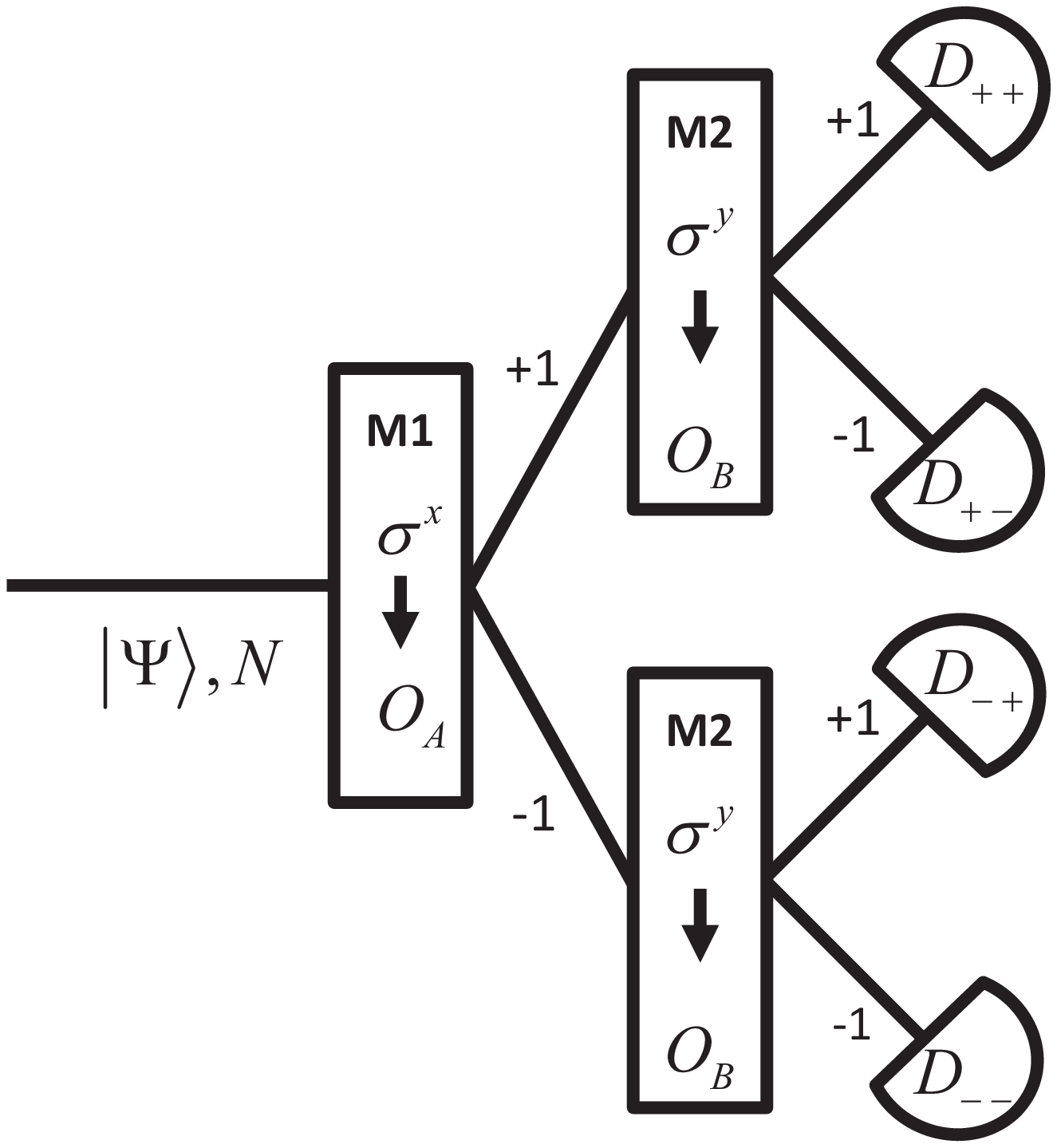}
\includegraphics[width=10cm]{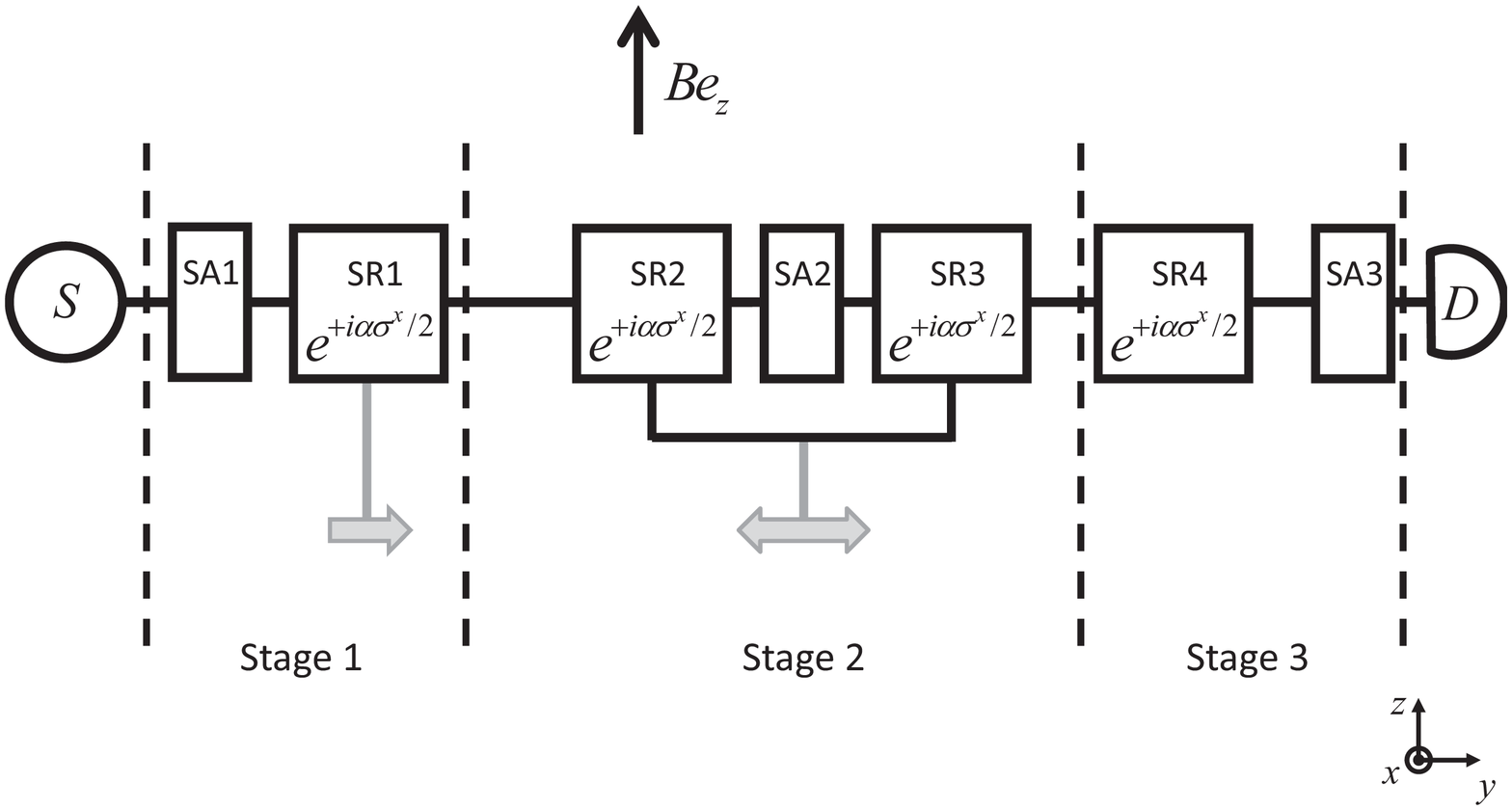}
\caption{%
Left: Conceptual experiment to validate Eq.~(\ref{ozawa1}).
The input consists of neutrons prepared in a state $|\psi\rangle$.
Apparatuses M1 and M2 perform Stern-Gerlach-like
selections of the neutrons based on the direction of their magnetic moments.
A successive measurement by M1 and M2 results in a click by one of the four detectors corresponding to the
four possible outcomes of the experiment. Repeated measurement for $N$ neutrons allows for a quantitative
determination of the error $\epsilon(A)$ and the disturbance $\eta (B)$ (see text).
Right: Block diagram of the laboratory experimental setup, see also Fig.~2 of \cite{ERHA12}. S: source;
SA1, SA2, SA3: spin analyzers; SR1, $\ldots$, SR4: spin rotators, rotating the magnetic moment of the neutron
by an angle $\alpha$ about the $x$ axis; D: detector; $B{\mathbf e}_z$: guiding magnetic
field along the $z$-direction. The positions of SR1 and of the pair (SR2,SR3) are variable.
}
\label{fig6}
\end{center}
\end{figure}
Figure \ref{fig6} (left) shows the conceptual experiment to determine the error and disturbance in the neutron experiment.
The error $\epsilon (A)$ is determined by data from apparatus $M1$
carrying out the projective measurement of $O_A$, and the disturbance $\eta (B)$ is determined by the data
from apparatus $M2$ carrying out the projective measurement of $B$ on the neutron spin state after the
measurement by $M1$. A succesive measurement of $O_A$ and $B$ has four possible outcomes. The outcome is registered
by one of the four detectors $D_{++}$, $D_{+-}$, $D_{-+}$ and $D_{--}$, where the first (second)
subscript refers to the two possible outcomes $\pm 1$ of apparatus $M1$ ($M2$).
Repeating the single neutron measurement $N$ times results in the counts $N_{++}$, $N_{+-}$, $N_{-+}$ and $N_{--}$ on
the respective detectors with $N=N_{++}+N_{+-}+N_{-+}+N_{--}$. From these counts the error $\epsilon(A)$
and disturbance $\eta (B)$ can be determined quantitatively~\cite{ERHA12}.
Apparatus $M1$ is performing a Stern-Gerlach type of measurement and is described by the measurement operator
$E_{\phi}(S_1)=(1+S_1\sigma^{\phi})/2$ with $S_1=\pm 1$ and
$O_A=\sum_{S_1=\pm 1}S_1E_{\phi}(S_1)$ so that $\langle\psi|O_A|\psi\rangle=(N_{++}+N_{+-}-N_{-+}-N_{--})/N$.
The modified measurement operator of apparatus $M2$ for the initial state of $M1$ is
$O_B=\sum_{S_2=\pm 1}E_{\phi}(S_2)BE_{\phi}(S_2)$. Hence, $\langle\psi|O_B|\psi\rangle=(N_{++}+N_{-+}-N_{+-}-N_{--})/N$.
Computation of $\epsilon (A)$ and $\eta (B)$ requires four different experiments with initial states
$|\psi\rangle=|z\rangle$, $|\psi\rangle=|-z\rangle$, $|\psi\rangle=|x\rangle$ and $|\psi\rangle=|y\rangle$
(see Eq.~(\ref{ozawa3})).

The block diagram of the laboratory single-neutron experiment is shown in Fig.~\ref{fig6} (right), see also Fig.~2 of
\cite{ERHA12}.
In stage 1, spin analyzer SA1 selects neutrons with spin up, that is with a magnetic moment pointing in the direction of the guiding magnetic field.
Spin rotator SR1 rotates the magnetic moment of the neutron by an angle $\alpha$ about the $x$ axis.
After SR1 the spin performs a Larmor precession about the $+z$ axis (due to the static magnetic guiding field).
By varying the position of SR1 arbitrary initial states (up to an irrelevant phase factor) can be produced at the end of stage 1.
The state $|+z\rangle$ is produced by switching off SR1, the state $|-z\rangle$ by setting
$\alpha =\pi$, the state $|+y\rangle$ by setting $\alpha =\pi/2$, the state $|+x\rangle$ by
setting $\alpha =\pi/2$ and changing the position of SR1 by one-quarter of the Larmor rotation period.
In stage 2, corresponding to measurement apparatus M1, the spin of the neutron in the prepared state performs a Larmor precession about the $+z$ axis.
By properly placing spin rotator SR2, that is by properly chosing the time that the spin performs Larmor precessions,
the spin component to be measured can be projected towards the $+z$ direction.
After passing SA2 in the $|+z\rangle$ state, SR3 produces the $|\pm \phi>$ eigenstate of $\sigma^{\phi}$
so that M1 makes the projective measurement of $O_A$.
In a similar way M2 measures B on the eigenstate $|\pm \phi\rangle$ in stage 3.

An important difference between the conceptual and laboratory experiment is that the spin analyzers select neutrons based on
the directions of their magnetic moments. Hence, one has to perform different experiments for the four different
settings $(S_1=\pm 1, S_2=\pm 1)$ of spin analyzers (SA2, SA3), which results in sixteen different experiments in total.
Also note that the number of input neutrons $N$ for each of those experiment is different from the number of detector clicks.

\subsection{Event-based model and simulation results}
\begin{figure}[t]
\includegraphics[width=7.5cm]{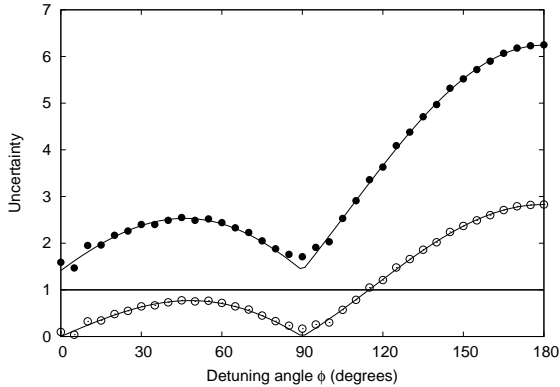}\hspace{2pc}%
\begin{minipage}[b]{7.5cm}
\caption{Simulation results for the uncertainties
$\epsilon (A)\eta (B) + \epsilon (A)\Delta (B) + \Delta (A)\eta (B)$
(solid circles) and $\epsilon (A)\eta (B)$ (open circles)
as a function of the detuning angle $\phi$ as obtained
from the event-by-event simulations of the single-neutron experiment~\cite{ERHA12,SULY13}.
Simulation parameters: $N=10000$.
Lines through the data points: quantum theoretical prediction as obtained from Eq.~(\ref{ozawa2}).
Horizontal line: lowerbound in Eq.~(\ref{ozawa2}).
}
\label{fig7}
\end{minipage}
\end{figure}


A minimal, discrete-event simulation model of the single-neutron experiment depicted in Fig.~\ref{fig6} (right)
requires a specification of the information carried by the neutrons, of the algorithm that simulates
the devices used in the experimental setup, and of the procedure to analyze the data.
All ingredients for simulating this experiment have already been described in the previous two sections.
Note that in this case no use of DLMs is required, since only simple spin manipulations (rotations and selections)
are involved.

For each pair of settings $(S_1,S_2)$ of the spin analyzers (SA2, SA3) and each position of the pair of spin
rotators (SR2, SR3) represened by a rotation of $\phi$ about the $z$ axis, the source sends $N$ neutrons
through the network (see Fig.~\ref{fig6} (right)). The source only sends a new neutron if the previous one has been counted
and destroyed by the detector or if the previous one was destroyed by one of the spin analyzers.
Counting the detection events allows for the calculation of $\epsilon ^2(A)$ and $\eta ^2(B)$.
The simulation results are presented in Fig.~\ref{fig7}.
It is clear that the simulation results are in excellent agreement with the quantum theoretical prediction Eq.~(\ref{ozawa2})
and with experimental results~\cite{ERHA12,SULY13} (not shown).

In conclusion, the event-based simulation approach produces the same results as the single neutron experiment
for the error of a spin-component measurement and the disturbance caused on another spin-component and
therefore also obeys the error-disturbance relation Eq.~(\ref{ozawa1}), derived using the theory of general quantum measurements.
In the event-based model no use is made of any
concept of quantum theory. The ``uncertainty'' in the event-based model comes from the spin analyzers
which transform time series of magnetic moments into random time series of $+1$'s and $-1$'s. This causes a loss of
``classical'' certainty. The event-based simulations of the neutron experiment thus demonstrate that the uncertainty relations are not
necessarily a signature of quantum physics. This is completely in agreement with the fact that for real data sets, uncertainty relations
provide theoretical bounds on the statistical uncertainties in the data.
Hence, the relevance of ``quantum theoretical'' uncertainty relations to laboratory experiments yielding real data needs to be scrutinized.

\section{Conclusion}
We have presented an event-based simulation method which allows for a mystery-free, particle-only description of interference and entanglement
phenomena observed in single-neutron interferometry experiments, and of a single neutron experiment designed to test the validity of Ozawa's
error-disturbance relation, a ``quantum theoretical'' uncertainty relation.
The method provides an ``explanation'' and ``understanding'' of what is going on in terms of elementary events, logic and arithmetic.
Note that a cause-and-effect simulation on a digital computer is a ``controlled  experiment'' on a macroscopic device which is logically equivalent to a mechanical device.
Hence, an event-by-event simulation that produces results which are usually thought of being of quantum mechanical origin, but emerge from a time series of discrete events
generated by causal adaptive systems, shows that there exists a macroscopic, mechanical model that mimics the underlying physical phenomena.

\ack
We would like to thank Helmut Rauch for many stimulating discussions.

\section*{References}
\bibliographystyle{iopart-num}
\bibliography{../../all13}

\providecommand{\newblock}{}
\begin{thebibliography}{10}
\expandafter\ifx\csname url\endcsname\relax
  \def\url#1{{\tt #1}}\fi
\expandafter\ifx\csname urlprefix\endcsname\relax\def\urlprefix{URL }\fi
\providecommand{\eprint}[2][]{\url{#2}}

\bibitem{HOME97}
Home D 1997 {\em {Conceptual Foundations of Quantum Physics}\/} (New York:
  Plenum Press)

\bibitem{BALL03}
Ballentine L~E 2003 {\em {Quantum Mechanics: A Modern Development}\/}
  (Singapore: World Scientific)

\bibitem{MICH11a}
{Michielsen} K, Jin F and {De Raedt} H 2011 {\em J. Comp. Theor. Nanosci.\/}
  {\bf 8} 1052 -- 1080

\bibitem{RAED12a}
{De Raedt} H and {Michielsen} K 2012 {\em Ann. Phys. (Berlin)\/} {\bf 524} 393
  -- 410

\bibitem{RAED12b}
{De Raedt} H, Jin F and {Michielsen} K 2012 {\em Quantum Matter\/} {\bf 1} 1 --
  21

\bibitem{RAUC74a}
Rauch H, Treimer W and Bonse U 1974 {\em Phys. Lett. A\/} {\bf 47} 369 -- 371

\bibitem{KROU00}
Kroupa G, Bruckner G, Bolik O, Zawisky M, Hainbuchner M, Badurek G, Buchelt
  R~J, Schricker A and Rauch H 2000 {\em Nucl. Instrum. Methods Phys. Res.
  A.\/} {\bf 440} 604 -- 608

\bibitem{HASE03}
Hasegawa Y, Loidl R, Badurek G, Baron M and Rauch H 2003 {\em Nature\/} {\bf
  425} 45 -- 48

\bibitem{ERHA12}
Erhart J, Sponar S, Sulyok G, Badurek G, Ozawa M and Hasegawa Y 2012 {\em Nat.
  Phys.\/} {\bf 8} 185 -- 189

\bibitem{SULY13}
Sulyok G, Sponar S, Erhart J, Badurek G, Ozawa M and Hasegawa Y 2013 {\em Phys.
  Rev. A\/} {\bf 88} 022110

\bibitem{FEYN65}
Feynman R~P, Leighton R~B and Sands M 1965 {\em The Feynman Lectures on
  Physics, Vol. 3\/} (Reading MA: Addison-Wesley)

\bibitem{YOUN02}
Young T 1802 {\em Phil. Trans. R. Soc. Lond.\/} {\bf 92} 12

\bibitem{DONA73}
Donati O, Missiroli G~P and Pozzi G 1973 {\em Am. J. Phys.\/} {\bf 41} 639 --
  644

\bibitem{MERL76}
Merli P~G, Missiroli G~F and Pozzi G 1976 {\em Am. J. Phys.\/} {\bf 44} 306 --
  307

\bibitem{TONO89}
Tonomura A, Endo J, Matsuda T, Kawasaki T and Ezawa H 1989 {\em Am. J. Phys.\/}
  {\bf 57} 117 -- 120

\bibitem{FRAB08}
Frabboni S, Gazzadi G~C and Pozzi G 2008 {\em Appl. Phys. Lett.\/} {\bf 93}
  073108

\bibitem{HASS10}
Hasselbach F 2010 {\em Rep. Prog. Phys.\/} {\bf 73} 016101

\bibitem{ROSA12}
{R Rosa} 2012 {\em Phys. Perspect.\/} {\bf 14} 178 -- 195

\bibitem{FRAB12}
Frabboni S, Gabrielli A, Gazzadi G~C, Giorgi F, Matteucci G, Pozzi G, Cesari
  N~S, Villa M and Zoccoli A 2012 {\em Ultramicroscopy\/} {\bf 116} 73 -- 76

\bibitem{BACH13}
Bach R, Pope D, Liu S and Batelaan H 2013 {\em New J. Phys.\/} {\bf 15} 033018

\bibitem{JACQ05}
Jacques V, Wu E, Toury T, Treussart F, Aspect A, Grangier P and Roch J~F 2005
  {\em Eur. Phys. J. D\/} {\bf 35} 561 -- 565

\bibitem{SAVE02}
Saveliev I~G, Sanz M and Garcia N 2002 {\em J. Opt. B: Quantum Semiclass.
  Opt.\/} {\bf 4} S477 -- S481

\bibitem{GARC02}
Garcia N, Saveliev I~G and Sharonov M 2002 {\em Phil. Trans. R. Soc. Lond. A\/}
  {\bf 360} 1039 -- 1059

\bibitem{KOLE13}
Kolenderski P, Scarcella C, Johnsen K, Hamel D, Holloway C, Shalm L, Tisa S,
  Tosi A, Resch K and Jennewein T Time-resolved double-slit experiment with
  entangled photons arXiv:1304.4943

\bibitem{ZEIL88}
Zeilinger A, G{\"{a}}hler R, Shull C~G, Treimer W and Mampe W 1988 {\em Rev.
  Mod. Phys.\/} {\bf 60} 1067 -- 1073

\bibitem{RAUC00}
Rauch H and Werner S~A 2000 {\em Neutron Interferometry: Lessons in
  Experimental Quantum Mechanics\/} (London: Clarendon)

\bibitem{KEIT91}
Keith D~W, Ekstrom C~R, Turchette Q~A and Pritchard D~E 1991 {\em Phys. Rev.
  Lett.\/} {\bf 66} 2693 -- 2696

\bibitem{CARN91}
Carnal O and Mlynek J 1991 {\em Phys. Rev. Lett.\/} {\bf 66} 2689 -- 2692

\bibitem{ARND99}
Arndt M, Nairz O, Vos-Andreae J, Keller C, {van der Zouw} G and Zeilinger A
  1999 {\em Nature\/} {\bf 401} 680 -- 682

\bibitem{BREZ02}
Brezger B, Hackerm{\"{u}}ller L, Uttenthaler S, Petschinka J, Arndt M and
  Zeilinger A 2002 {\em Phys. Rev. Lett.\/} {\bf 88} 100404

\bibitem{JUFF12}
Juffmann T, Milic A, {M M\"ullneritsch}, Asenbaum P, Tsukernik A, {T\"uxen} J,
  Mayor M, Cheshnovsky O and Arndt M 2012 {\em Nature Nanotechnology\/} {\bf 7}
  297 -- 300

\bibitem{COUD06}
Couder Y and Fort E 2006 {\em Phys. Rev. Lett.\/} {\bf 97} 154101

\bibitem{HASE11}
Hasegawa Y and Rauch H 2011 {\em New J. Phys.\/} {\bf 13} 115010

\bibitem{LEMM10}
Lemmel H and Wagh A~G 2010 {\em Phys. Rev. A\/} {\bf 82} 033626

\bibitem{NIEU13}
Allahverdyan A~E, Balian R and Nieuwenhuizen T~M 2013 {\em Phys. Rep.\/} {\bf
  525} 1 -- 166

\bibitem{BORN64}
Born M and Wolf E 1964 {\em {Principles of Optics}\/} (Oxford: Pergamon)

\bibitem{COUD10}
Fort E, Eddi A, Boudaoud A, Moukhtar J and Couder Y 2010 {\em PNAS\/} {\bf 107}
  17515--17520

\bibitem{COUD11}
Eddi A, Sulta E, Moukhtar J, Fort E, Rossi M and Couder Y 2011 {\em J. Fluid
  Mechan.\/} {\bf 674} 433--463

\bibitem{COUD12}
Couder Y and Fort E 2012 {\em J. Phys.: Conf. Ser.\/} {\bf 361} 012001

\bibitem{JIN10b}
{Jin} F, {Yuan} S, {De Raedt} H, {Michielsen} K and Miyashita S 2010 {\em J.
  Phys. Soc. Jpn.\/} {\bf 79} 074401

\bibitem{RAED05b}
{De Raedt} K, {De Raedt} H and Michielsen K 2005 {\em Comp. Phys. Comm.\/} {\bf
  171} 19 -- 39

\bibitem{RAED05d}
{De Raedt} H, {De Raedt} K and Michielsen K 2005 {\em Europhys. Lett.\/} {\bf
  69} 861 -- 867

\bibitem{BART09}
Bartsch C and Gemmer J 2009 {\em Phys. Rev. Lett\/} {\bf 102} 110403

\bibitem{BOHM51}
Bohm D 1951 {\em Quantum Theory\/} (New York: Prentice-Hall)

\bibitem{WEIH98}
Weihs G, Jennewein T, Simon C, Weinfurther H and Zeilinger A 1998 {\em Phys.
  Rev. Lett.\/} {\bf 81} 5039 -- 5043

\bibitem{WEIH00}
Weihs G 2000 {\em {Ein Experiment zum Test der Bellschen Ungleichung unter
  Einsteinscher Lokalit\"at}\/} Ph.D. thesis University of Vienna
  {\url{http://www.uibk.ac.at/exphys/photonik/people/gwdiss.pdf}}

\bibitem{HNIL02}
Hnilo A, Peuriot A and Santiago G 2002 {\em Found. Phys. Lett.\/} {\bf 15} 359
  -- 371

\bibitem{AGUE09}
{Ag\"uero} M~B, Hnilo A~A, Kovalsksy M~G and Larotonda M~A 2009 {\em Eur. Phys.
  J. D\/} {\bf 55} 705 --709

\bibitem{CLAU69}
Clauser J~F, Horne M~A, Shimony A and Holt R~A 1969 {\em Phys. Rev. Lett.\/}
  {\bf 23} 880 -- 884

\bibitem{BASU01}
Basu S, Bandyopadhyay S, Kar G and Home D 2001 {\em Phys. Lett. A\/} {\bf 279}
  281 -- 286

\bibitem{BELL93}
Bell J~S 1993 {\em {Speakable and Unspeakable in Quantum Mechanics}\/}
  (Cambridge: Cambridge University Press)

\bibitem{CIRE80}
Cirel'son B~S 1980 {\em Lett. Math. Phys.\/} {\bf 4} 93 -- 100

\bibitem{Bartosik2009}
Bartosik H, Klepp J, Schmitzer C, Sponar S, Cabello A, Rauch H and Hasegawa Y
  2009 {\em Phys. Rev. Lett.\/} {\bf 103} 040403

\bibitem{OZAW03}
Ozawa M 2003 {\em Phys. Rev. A\/} {\bf 67} 042105

\bibitem{FUJI12}
Fujikawa K 2012 {\em Phys. Rev. A\/} {\bf 85}(6) 062117

\bibitem{FUJI13a}
Fujikawa K and Umetsu K 2013 {\em Prog. Theor. Exp. Phys.\/} {\bf 2013} 013A03

\bibitem{HEIS27}
Heisenberg W 1927 {\em Z. Phys.\/} {\bf 43} 172--198

\bibitem{BUSH07}
Busch P, Heinonen T and Lahti P 2007 {\em Physics Reports\/} {\bf 452} 155 --
  176

\bibitem{BUSH13}
Busch P, Lahti P and Werner R~F 2013 {\em Phys. Rev. Lett.\/} {\bf 111} 160405

\bibitem{BALL70}
Ballentine L~E 1970 {\em Rev. Mod. Phys.\/} {\bf 42}(4) 358 -- 381

\end{thebibliography}
\end{document}